# Hardware-aware Lightweight Photonic Spiking Neural Network for Pattern Classification

Shuiying Xiang[1]*, Yahui Zhang[1], Shangxuan Shi[1], Haowen Zhao[1], Dianzhuang Zheng[1], Xingxing Guo[1], Yanan Han[1], Ye Tian[1], Liyue Zhang[2], Yuechun Shi[3], & Yue Hao[1]

[1]State Key Laboratory of Integrated Service Networks, State Key Discipline Laboratory of Wide Bandgap Semiconductor Technology, Xidian University, Xi'an 710071, China;

[2]Key Laboratory of Photonic-Electronic Integration and Communication-Sensing Convergence (Ministry of Education), Southwest Jiaotong University, Sichuan, 611756, China;

[3]Yongjiang laboratory, No. 1792 Cihai South Road, Ningbo 315202, China.

*Corresponding author: syxiang@xidian.edu.cn

There exists a significant scale gap between photonic neural network integrated chips and neural networks, which hinders the deployment and application of photonic neural network. Here, we propose hardware-aware lightweight spiking neural networks (SNNs) architecture tailored to our photonic neuromorphic chips, and conducts hardware-software collaborative computing for solving patter classification tasks. Here, we employed a simplified Mach-Zehnder interferometer (MZI) mesh for performing linear computation, and 16-channel distributed feedback lasers with saturable absorber (DFB-SA) array for performing nonlinear spike activation. Both photonic neuromorphic chips based on the MZI mesh and DFB-SA array were designed, optimized and fabricated. Furthermore, we propose a lightweight spiking neural network (SNN) with discrete cosine transform to reduce input dimension and match the input/output ports number of the photonic neuromorphic chips. We demonstrated an end-to-end inference of an entire layer of the lightweight photonic SNN. The hardware-software collaborative inference accuracy is 90% and 80.5% for MNIST and Fashion-MNIST datasets, respectively. The energy efficiency is 1.39 TOPS/W for the MZI mesh, and is 987.65 GOPS/W for the DFB-SA array. The lightweight architecture and experimental demonstration address the challenge of scale mismatch between the photonic chip and SNN, paving the way for the hardware deployment of photonic SNNs.

## 1. INTRODUCTION

With the rapid development of artificial intelligence (AI) represented by ChatGPT and Deepseek, the demand for computing power has shown explosive growth. Traditional electronic computing chips based on the von Neumann architecture are facing severe challenges of the "memory wall" and "power wall", and their computing power growth is gradually approaching physical limits. Photonic neural networks (PNNs) stand out with their ultra-high speed, ultra-low power consumption, and inherent large-scale parallel processing capabilities, becoming one of the most promising disruptive computing paradigms in the post-Moore era [1-12].

In recent years, PNNs based on Mach-Zehnder interferometers (MZI) [13-23], micro-ring resonator [24-27], phase change material crossbars [28-30], and spiking laser neurons [31-37] have been studied extensively. Specifically, programmable MZI mesh-based PNN chips have been developed rapidly since 2017, with their matrix scale gradually advancing from the initial 4×4 [13, 14] to the sophisticated 128×128 [22, 23]. For diverse photonic computing paradigms, reported PNNs implementations cover photonic multi-layer perceptrons [38-39], convolutional neural networks [40-43], tensor processing [44-48], and diffractive neural networks [49-55].

Among numerous neural network models, spiking neural networks (SNNs) have attracted significant attention due to their adoption of the efficient and sparse processing mechanisms of the biological brain. Combining the ultra-high speed and low-power advantages of photonic technology with the biologically inspired efficiency of SNNs has spawned the emerging interdisciplinary field of photonic spiking neural networks (PSNNs). PSNNs are expected to achieve remarkable computing speeds while keeping energy consumption at an extremely low level, providing an ideal hardware solution for scenarios such as edge computing and real-time intelligent processing. In addition, these photonic SNN systems exhibit outstanding noise immunity while alleviating the stringent requirements for high-precision analog-to-digital (AD) and digital-to-analog (DA) conversions during the input of spike signals and retrieval of spike output data. Consequently, PSNN emerges as a hardware-friendly architecture. Recently, PSNNs have attracted numerous attention [28, 32-37, 56-60]. However, there is a huge gap between the scale of actual photonic neuromorphic chips and the scale of SNNs required to complete specific tasks. This mismatch in scale between "software" (neural network models) and "hardware" (neuromorphic chips) seriously hinders the transition of PSNNs from theoretical models to practical applications.

To address this core contradiction, the key lies in model compression of PSNNs. That is, on the premise of maintaining their core functions and performance, the complexity and scale of the model are significantly reduced to enable efficient deployment on resource-constrained photonic chips. Traditional model compression methods, such as pruning and quantization, have been applied in electronic SNNs, but they often fail to fully consider the unique physical characteristics and hardware constraints of photonic computing. Therefore, there is an urgent need for a new lightweight architecture that is highly compatible with the characteristics of photonic hardware.

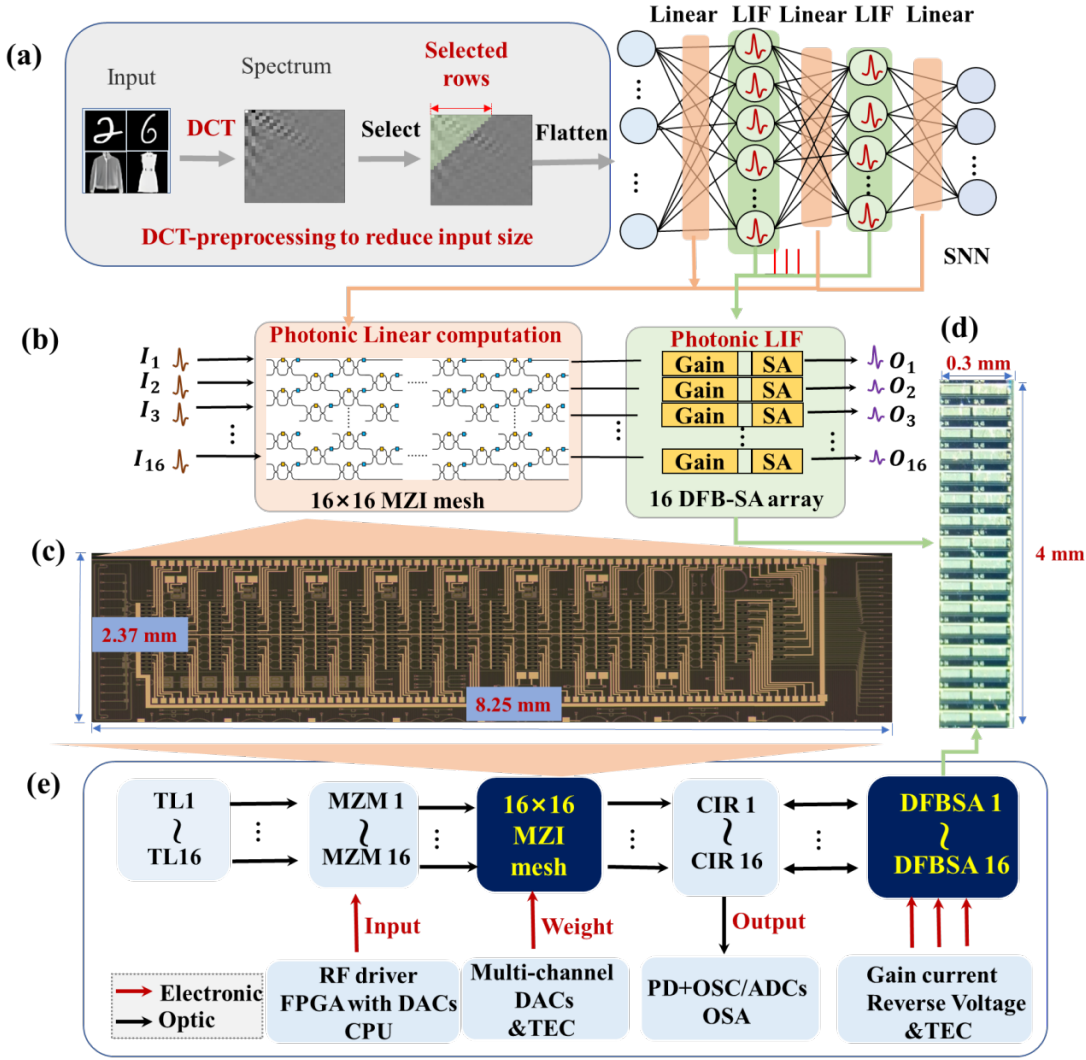

Fig. 1. (a) The overall architecture of the frequency-domain SNN with DCT, (b) schematic diagram of photonic linear computation and photonic LIF neuron, (c) microscopic image of the fabricated MZI mesh chip, (d) microscopic image of the fabricated DFB-SA array chip (e) The testing experimental setup.

We propose a lightweight PSNN architecture based on discrete cosine transform (DCT) [61], and deploys this network on a photonic neuromorphic chip for hardware-software collaborative computing, aiming to fundamentally bridge the gap between algorithm scale and hardware chip size. We introduce DCT into the design of PSNNs to reduce the dimensionality and remove redundancy of input data, retaining only the most informative frequency components as inputs to the SNN. This fundamentally reduces the dimensionality of processed data and the scale required for subsequent networks, thereby making it possible to deploy the entire SNN on a photonic chip. We consider a fully connected network architecture in order to achieve single-time-step SNN training and completed classification tasks on MNIST and Fashion-MNIST datasets. The hardware-software inference based on the photonic neuromorphic chip and the lightweight PSNN are further carried out. The work not only provides an innovative technical solution to address the scale bottleneck of PSNNs, but also takes a key step in promoting the transition of photonic neuromorphic computing from the laboratory to practical applications such as edge computation and embodied intelligence.

## 2. Experimental setup and Method

A. Lightweight architecture of PSNN with DCT.

To match the limited fan-in size of the photonic neuromorphic computing chip, we propose a frequency-domain pruning method by applying DCT to the input image of the SNN. As shown in Fig.1(a), after DCT, the frequency spectrum of the input image is obtained. We select the low-frequency components and feed them into the SNN for further processing, which significantly reduces the input size of the SNN. The frequency-domain components are flattened and then fed into the first linear layer of the SNN. Following the linear layer, the weighted summation signals are conveyed to the leaky integrate-and-fire (LIF) neuron for nonlinear spike activation. Different hidden layers can be considered for different tasks.

The DCT is a mathematical transformation that converts spatial domain data into the frequency domain, capturing the frequency characteristics of the input data. In image processing, two-dimensional (2D) DCT is applied to transform a spatial-domain image (e.g., a 28×28 MNIST image) into its frequency spectrum, where the coefficients represent the energy distribution at different frequency components. Mathematically, the 2D DCT is defined as:

$$F(u,v) = \sum_{x=0}^{N-1}\sum_{y=0}^{N-1} I(x,y) \cos\left[\frac{\pi(2x+1)u}{2N}\right] \cos\left[\frac{\pi(2y+1)v}{2N}\right],$$

where $I(x,y)$ is the input image pixel value, and $F(u,v)$ represents the frequency coefficients after transformation. $u, v$ are the indices of the frequency components, and $N$ is the image dimension.

The 2D DCT concentrates low-frequency components in the top-left corner of the transformed spectrum, while the high-frequency components spread towards the bottom-right corner. In our scheme, the upper triangular region of the 2D DCT spectrum is extracted, retaining only the most significant low-frequency features. This reduces the input data dimension while preserving essential information, thereby simplifying the computational complexity for the subsequent neural network processing.

### B. Photonic synapse chip based on simplified MZI mesh.

As shown in Fig. 1(b), we deployed the linear layer on a photonic synapse array chip to perform photonic linear computation. For this purpose, we designed and fabricated a chip based on a simplified MZI mesh on a silicon photonic platform to execute incoherent optical matrix-vector multiplication (MVM). The structure with a 16 × 16 MZI mesh are designed, where each MZI incorporates only a single phase shifter on one of its inner arms. For representing a 16 × 16 weight matrix, the design utilizes 17 × (17-1)/2 + 16 = 152 phase shifters. Thus, this structure achieves significant area, transmission loss and power consumption reduction. The chip features a length of 8.25 mm, a width of 2.37 mm, and a corresponding area of 19.55 mm², with its microscopic image presented in Fig. 1(c). When light is injected through a single port, the chip demonstrates an approximate insertion loss of 13 decibels (dB). Equipped with 152 phase shifters (each featuring a half-wave phase shift power $P_\pi$=30mW), the total power consumption is about 152×0.03W/2=2.28 W.

### C. Photonic nonlinear spiking neuron chip based on DFB-SA.

As shown in Fig. 1(b), the LIF layer was deployed on a 16-channel DFB-SA laser array, fabricated on a III-V platform to enable element-wise nonlinear spike activation. As presented in Fig. 1(d), the array chip measures 4 mm in length and 0.3 mm in width, with a total area of 1.2 mm². Each single-channel DFB-SA laser within the array is structured with two sections, namely a gain region and a saturable absorber (SA) region. The gain region is forward-biased by a current source ($I_G$), while the saturable absorber (SA) region is reverse-biased by a voltage source ($V_{SA}$). The epitaxial wafer structure for single-channel DFB-SA laser unit is presented in Fig. 2(a). The optimized multi-quantum well (MQW) structure comprises 7 well layers, and the thickness of the upper and lower confining layers is approximately 70 nm. Furthermore, anti-reflection (AR) coatings and high-reflection (HR) coatings are deposited on the two laser facets to boost the light emission power, while the SA region is placed close to the HR side. As shown in Fig. 2(b), the measured lasing threshold is about 28 mA, and is almost the same for all the 16 DFB-SA lasers. The self-pulsation outputs, the corresponding frequency spectra and optical spectra for different $I_G$ and $V_{SA}$ are displayed in Fig. 2(c). The maximum self-pulsation frequency is measured to be about 5 GHz. When biasing slightly below the self-pulsation state, the nonlinear neuron-like response can be achieved [28], similar to our previous demonstration of single DFB-SA laser. The optical spectrum of the DFB-SA laser broadens when the device operates in the self-pulsation state.

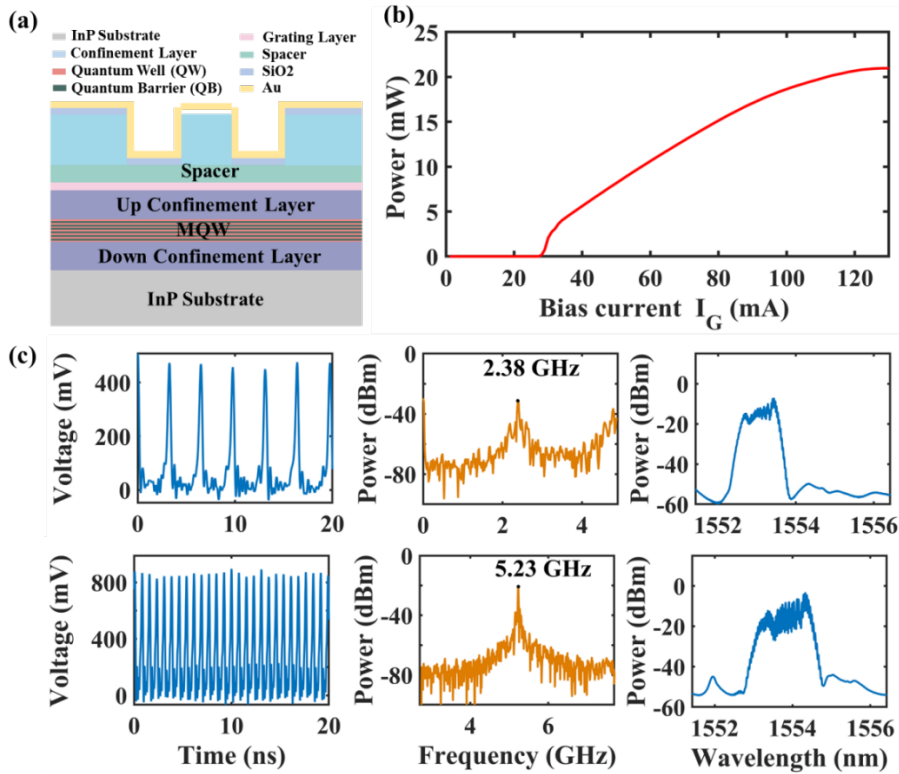

Fig. 2. The epitaxial wafer structure and properties of DFB-SA laser. (a) The schematic of epitaxial wafer structure of DFB-SA laser. (b) Power current curve of DFB-SA laser. (c) The self-pulsation, frequency spectra of self-pulsation state and optical spectra of DFB-SA laser.

D. Experimental setup for testing the photonic neuromorphic computing system.

The experimental setup of photonic neuromorphic computing system for deploying frequency-domain SNN is illustrated in Fig. 1 (e). Optical carriers with different wavelengths are generated by multi-channel TLs. The input signals for the photonic neuromorphic computing system were generated by a Xilinx FPGA platform, ZCU216 evaluation board, which features a Zynq UltraScale+ RFSoC 49DR chip and 16-channel high-speed AD/DA converters. The FPGA was controlled by a digital computer. The input signals were modulated into optical carriers by the Mach-Zehnder modulator (MZM). The modulated optical signal was then injected into the MZI mesh. The weights of the frequency-domain SNN were deployed on the MZI mesh, which was controlled through a multi-channel voltage source and a TEC for thermal stabilization. The modulated optical input signal was multiplied and added in the MZI mesh chip. To perform the nonlinear computation, the 16-channel output from the MZI mesh was fed into a 16-channel DFB-SA laser array. The output from the 16-channel DFB-SA laser array was routed through an array of three-port optical circulators. The resulting optical signals were then characterized using an optical spectrum analyzer and converted into electrical signals by a photodetector (Agilent/HP 11982A) for subsequent acquisition with an oscilloscope (Keysight DSOZ592A).

## 3. Results

A. Software-hardware collaborative training-inference framework.

To tackle the challenges of SNN training difficulty and accuracy degradation in optical computing, we propose an end-to-end three-stage software-hardware-software collaborative training-inference framework tailored for frequency-domain photonic SNNs. Three sequential phases are incorporated into this framework, namely software pre-training phase, local photonic hardware in-situ training phase, and hardware-aware software fine-tuning phase. The complete workflow of the hardware-software integrated training and inference process is illustrated in Fig. 3.

**SNN pre-training.** As shown in Fig. 3(a), the process begins with training an ANN, which is then converted to a SNN. Next, direct training of the SNN is performed using backpropagation based on surrogate gradients. Notably, a temporal pruning method is adopted to achieve low latency inference. This method treats the threshold and leakage of SNN as trainable parameters, initiating training with an SNN configured for T time steps and progressively reducing the time steps in each iteration, ultimately yielding an SNN with a single time step (T=1). In our implementation, the time steps of SNN are gradually compressed from 5 to 3, and finally to 1. We designate the software-trained weights requiring mapping to the hardware MZI chip as $W^S$. The complete hybrid training algorithm, incorporating DCT, ANN-to-SNN conversion, direct training with surrogate gradient and temporal pruning, is summarized in *Algorithm 1*.

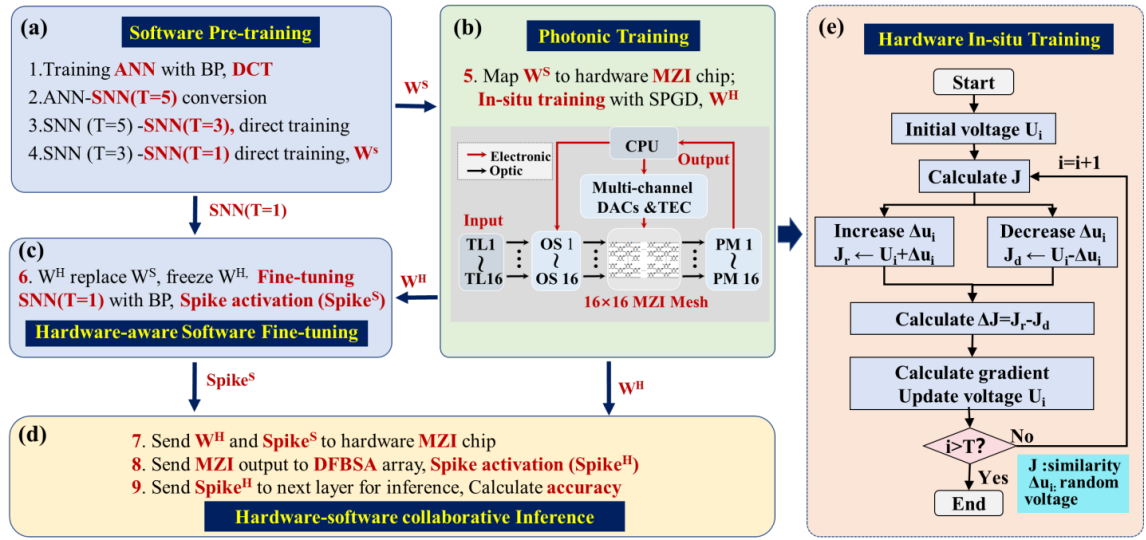

Fig. 3. software-hardware-software collaborative training-inference framework for frequency-domain photonics SNNs.

**Algorithms 1. Single time step SNN training with DCT and temporal pruning**

**Input:** Input 28×28 MNIST image $I$, number of selected rows in spectrum diagram $N_r$, trained the SNN model with N time steps ($T_N$), the time steps reduction step size $b$, number of epochs to train $e$.

**DCT Pre-processing:**
Apply DCT to $I$ to obtain a 28×28 spectrum diagram $F$.
Extract the upper triangle region from the spectrum diagram using $N_r$:
**for** each pixel $F(i,j)$:
    Include pixel if $i + j \leq N_r - 1$
// The size of input data for the network is $\frac{N_r \times (N_r+1)}{2}$

**Training with temporal pruning:**
Train an ANN, new SNN initialized with trained parameters of $T_N$, reduced latency $T_r = N - b$
**while** $T_r \geq 1$ **do**
    // Training Phase
    **for** epoch 1: $e$ **do**
        // Train network with $T_r$ timesteps
    **end for**
    // Initialize another iso-architecture SNN with parameters of above trained network
    // Temporal pruning
    $T_r = T_r - b$
**end while**

**Local photonic hardware in-situ training.** To represent the target weight matrix $W^S$, the MZI mesh chip is configured using the stochastic parallel gradient descent (SPGD) algorithm based on local in-situ training [18]. This configuration strategy is due to the implicit nature of the optical matrix characterized by the MZI mesh. Specifically, there is no direct correspondence between one or more phase shifter heater voltages and a single element of the matrix. However, the weights trained on the hardware MZI chip may not perfectly match $W^S$, as they are affected by noise, fabrication imperfections, and the limited precision of the multi-channel DACs that drive the phase shifters. The weight matrix represented by the trained MZI mesh is denoted as $W^H$.

The experiment setup for configuring MZI mesh chip is depicted in Fig. 3(b). Continuous wave is generated by a 16-channel TL source. A 16-channel TL source generates the CW input, while a 16-channel optical switch (OS) controls the injection sequence into the input ports of the MZI chip. Multi-channel voltage sources, controlled by a host computer, are used to adjust the weighting of the MZI. The output power of MZI chip is detected by the 16-channel optical power meters (PM) and than is feedback to the host computer. For the iterative update of MZI mesh weights, the SPGD algorithm is implemented to optimize the driving voltage of each phase shifter within the MZI mesh. The optimized voltages are then loaded onto the corresponding phase shifters by the host computer through regulating the multi-channel voltage sources via an Ethernet link. To mitigate the influence of temperature fluctuations on device performance, the operating temperature of the MZI mesh chip is actively stabilized at 25°C by the TEC.

The progress of the in-situ training for the MZI chip is illustrated in Fig. 3(e), which primarily consists of system initialization, weight matrix characterization, performance evaluation, gradient estimation via perturbation, and iterative optimization five key steps.

①System initialization. A set of random voltages $U_i$ is applied to the MZI phase shifters to initialize the MZI mash chip.

②Weight matrix characterization. To characterize the weight matrix, CW light is first injected into a single input port. The optical power measured from the 16 output ports then forms one column of the matrix. This process is repeated by sequentially injecting light into each of the 16 input ports, ultimately constructing the complete 16×16 hardware weight matrix $W^{H1}$.

③ Performance evaluation. The similarity $J$ between the characterized hardware weight matrix $W^{H1}$ and the target matrix $W^S$ is quantified using cosine similarity [18].

④Gradient estimation. The gradient is estimated by applying positive and negative voltage random perturbations ($U_i+\Delta u_i$ and $U_i-\Delta u_i$), remeasuring the similarity metric ($J_r$ and $J_d$), and computing the difference $\Delta J = J_r - J_d$. The phase shifter voltages are then updated by calculating the gradient based on $\Delta J$ and $\Delta u_i$.

⑤Iterative optimization. The phase shifter voltages are updated based on the calculated gradients, and steps 2-5 are repeated until the hardware weight matrix converges to the target.

**Hardware-aware software fine-tuning stage.** As shown in Fig. 3(c), the software-trained weights $W^S$ are substituted with their hardware-calibrated counterparts $W^H$. During the subsequent fine-tuning stage, the weights corresponding to the MZI layer ($W^H$) are fixed, while the remaining weights in the network are adjusted. This hardware-aware fine-tuning strategy ensures that the weights of the MZI layer are accurately represented by the physical chip, effectively compensating for inherent manufacturing variations and limitation of hardware.

**Hardware-software collaborative inference.** As illustrated in Fig. 3(d), following the hardware-aware software fine-tuning stage, the converged weights $W^H$ and input spike signals are sent to the MZI chip for optical processing. The output of MZI is then fed into a DFBSA array to generate activated spikes (Spike$^H$). Spike$^H$ are subsequently transmitted to the next software-based layer for further inference, with the aim of calculating accuracy.

## B. MNIST and Fashion-MNIST classification using photonic neuromorphic chips.

We first consider a lightweight frequency-domain SNN with a size of 45×16×16×10 to classify the MNIST dataset. Specifically, after applying the DCT, we select only 45 low-frequency components for classification with frequency-domain SNN. The network comprises one input layer with 45 neurons, two hidden layers with 16 neurons each, and one output layer with 10 neurons. Such a small input layer size matches well with the fan-in number of an MZI mesh chip that can be fabricated using currently available silicon photonics process. In our experimental demonstration, the entire hidden layer with a size of 16×16 is deployed to the fabricated MZI mesh chip and the DFB-SA array chip.

The pre-training and hardware-aware fine-tuning results are presented in Fig. 4(a). Here, the evolution of the training and testing accuracy and loss, as well as the confusion matrix are shown. We can find that the accuracy is 95.66% for ANN, 94.4% for SNN with T=5, 94.04% for SNN with T=3, 91.23% for SNN with T=1, and reaches 90.17% after performing hardware-aware fine-tuning. The hardware in-situ training results are presented in Fig. 4(b). It can be found that, even with only a single time step and with the input size of network reduced to only 45 (compared to 784 for conventional scheme), the accuracy maintains a relatively high value. Thus, this lightweight frequency-domain SNN with DCT closely matches the size of available photonic SNN chips, effectively bridging the gap between the network size and the integrated photonic chip scale.

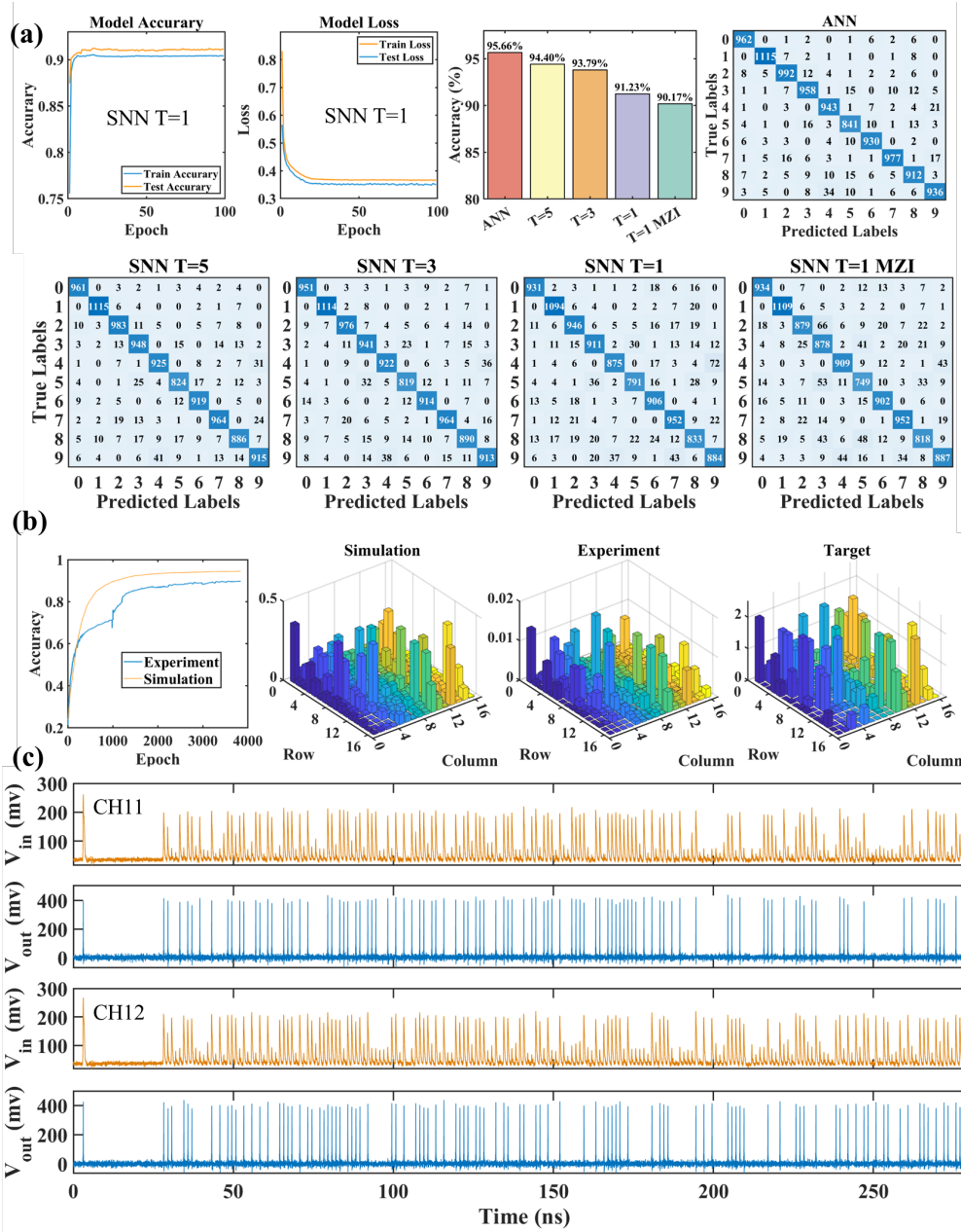

Fig. 4. The results for the training and inference stage for MNIST datasets. (a) Accuracy and loss as functions of epoch, the test accuracy and confusion matrices for ANN training, SNN with T=5, SNN with T=3, SNN with T=1, and hardware-aware fine-tuning SNN with T=1, (b) Local in-situ training results including experimental and simulated accuracy as functions of epoch. (c) the experimentally measured linear computation and nonlinear spike activation results of two representative channels.

By deploying the weights to the MZI mesh chip, we measured the hardware linear computation and nonlinear spike activation results for 200 randomly selected test images for hardware and software inference. The experimentally measured results are presented in Fig. 4(c). Here, the temporal results of two channels are displayed for simplicity. Clearly, the MZI mesh chip can perform linear MVM function, and the DFB-SA laser array can accurately perform the nonlinear spike activation. Based on the neuron-like response of the DFB-SA laser, input pulses with relatively small amplitudes cannot trigger spike responses, and the nonlinear outputs become much sparser than the linear outputs.

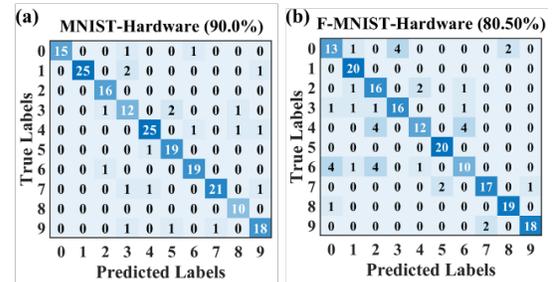

Fig.5. The experimentally measured testing accuracy for (a) MNIST and (b) Fashion-MNIST dataset.

The confusion matrix for the hardware chip inference for the MNIST dataset is presented in Fig. 5(a), and the hardware testing accuracy is 90%, which is slightly lower than 90.17% that achieved in the hardware-aware software fine-tuning stage, indicating that the proposed frequency-domain SNN is robust to fabrication errors and system noise.

We also use the same frequency-domain SNN network structure and photonic neuromorphic computing chips to classify the Fashion-MNIST. The results of training accuracy and loss, confusion matrices, as well as the photonic hardware in-situ training results can be found in Figs. 6 (a)-(b). We find that the accuracy is 85.57% for ANN, 84.45% for SNN with T=5, 84.35% for SNN with T=3, 81.23% for SNN with T=1, and 80.66% for hardware-aware fine-tuning SNN with T=1. The hardware linear and nonlinear computation temporal results for Fashion-MNIST can be found in Fig. 6 (c). Correspondingly, as shown in Fig.5 (b), the hardware testing accuracy is 80.50% for the Fashion-MNIST dataset, and is slightly lower than 80.66% for software inference.

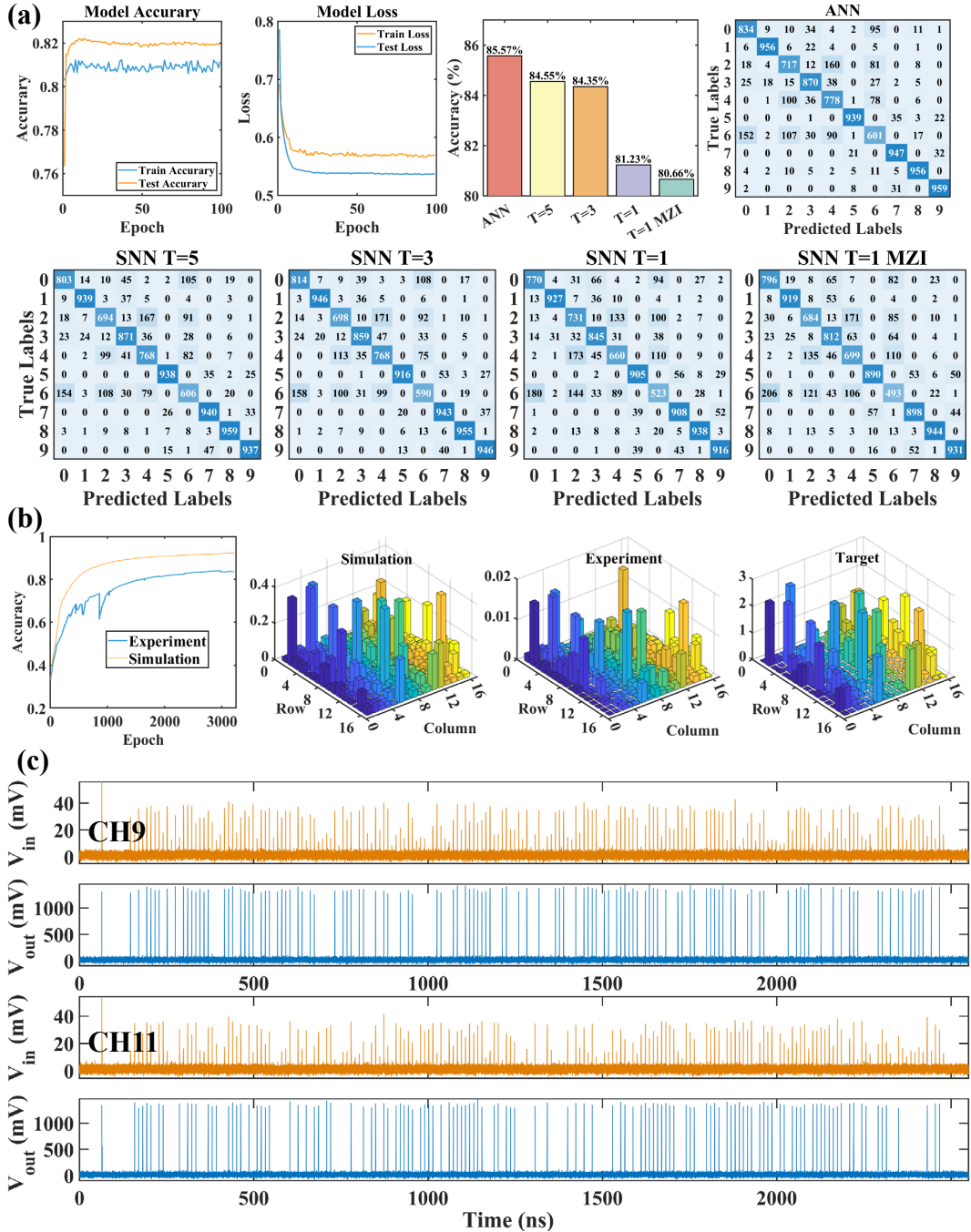

Fig.6. The results for the training and inference stage for Fashion-MNIST dataset. (a) Accuracy and loss as functions of epoch, the test accuracy and confusion matrices for ANN training, SNN with T=5, SNN with T=3, SNN with T=1, and hardware-aware fine-tuning SNN with T=1. (b) Local in-situ training results including experimental and simulated accuracy as functions of epoch. (c) the experimentally measured linear computation and nonlinear spike activation results of two representative channels.

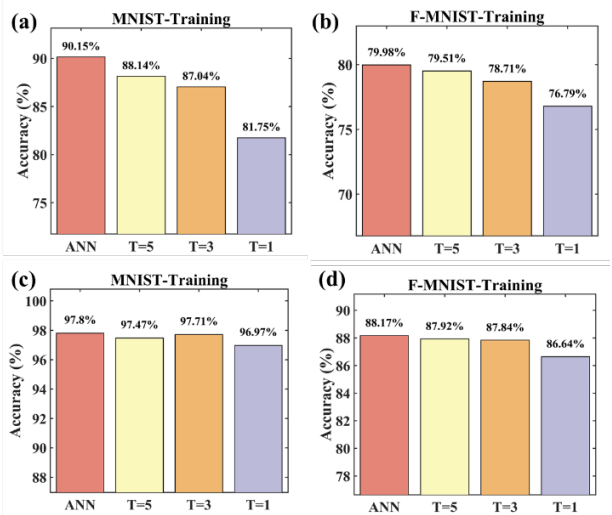

Fig.7. SNN pre-training results for (a) MNIST and (b) Fashion-MNIST dataset with a size of 15×16×16×10. SNN pre-training results for (c) MNIST and (d) Fashion-MNIST dataset with a size of 120×64×64×10.

We further consider numerical simulations to explore different network sizes. The numerical results for frequency-domain SNNs with size of 15×16×16×10 and 120×64×64×10 are presented for both datasets in Fig.7. For MNIST dataset, the accuracy for SNN with T=1 is 81.57% for the network size of 15×16×16×10, and is 96.68% for the network size of 120×64×64×10. For the Fashion-MNIST, the accuracy for SNN with T=1 is 76.79% for the network size of 15×16×16×10, and is 86.56% for the network size of for 120×64×64×10. The numerical findings validate the feasibility of enhancing the inference accuracy with larger-scale SNNs.

## 4. DISCUSSION AND CONCLUSION

**Metrics.** For PSNNs encompassing both linear and nonlinear computing, we evaluated the key performance metrics of the MZI mesh chip and the DFB-SA laser array chip, while accounting for the constraint relationship between them. Given that the DFB-SA laser array operates at the maximum self-pulsation frequency of 5 GHz, the signal rate of the MZI is correspondingly configured to 5 GHz. Under this operational parameter setting, the computing rate of the MZI chip is calculated as 2 × 16 × 16 × 5 GHz, yielding a value of 2.5 TOPS. Leveraging the metrics of the area and total power consumption of chip, we calculated its energy efficiency at 1.39 TOPS/W and computing density at 0.13 TOPS/mm². As for the 16-channel DFB-SA laser array chip, its throughput is calculated as 640 GOPS, considering that each input undergoes 8 equivalent operations when emulating the LIF neuron, with the formulation being 16×8×5 GHz. With a total energy consumption of approximately 0.648 W, the chip achieves an energy efficiency of 987.65 GOPS/W and a computing density of 533.33 GOPS/mm². The end-to-end latency for implementing a full layer of the SNN, including both photonic MVM and photonic LIF modules, is calculated as 320 ps. Table 1 presents a comparative analysis of our chips against state-of-the-art PNN chips including optical nonlinear computation and electronic neuromorphic chips. To the best of our knowledge, among previously reported PNN chips capable of supporting optical nonlinear computing, our design is superior to counterparts in energy efficiency, computing density, and the number of trainable parameters three critical dimensions.

**Scalability.** Notably, the photonic spiking neurons based on the DFB-SA laser array exhibit excellent scalability. They can be readily expanded to 150 channels while maintaining high wavelength precision [64]. For the MZI mesh architecture, prior work has already demonstrated the feasibility of large-scale integration. Specifically, a 128×128 integrated photonic arithmetic computing engine has been successfully implemented for optical Ising model applications [20]. Building on this precedent, our proposed simplified MZI mesh design enables scalable expansion to a 128×128 configuration with manageable loss levels. Furthermore, advanced heterogeneous integration technologies offer the potential to achieve even tighter integration of photonic synapse chips and photonic spiking neuron chips, laying the groundwork for higher-performance monolithic systems.

**All-optical DCT potential.** DCT encoding, a well-established and efficient image processing technique, can be inherently implemented in diffractive neural networks relying on optical passive devices [65]. Looking ahead, integration with this all-optical DCT paradigm could empower all-optical pattern classification to enable pattern preprocessing and recognition without the need for EO/OE conversions. This integration would leverage inherent optical signal processing capabilities of the system, achieving remarkable low latency and superior energy efficiency.

Table 1. Comparison with state-of-the-art photonic neural network chips and electronic neuromorphic chips

| Metrics | Computing density | Energy efficiency | Trainable parameters | Optical nonlinear computation | For SNN |
|---|---|---|---|---|---|
| **Our work** | 0.13 TOPS/mm² | 1.39 TOPS/W | **272** | √ | √ |
| Ashtiani et al [15] | 1.75 TOPS/mm² | 2.9 TOPS/W | 67 | √ | **X** |
| Bandyopadhyay et al [17] | 0.02 TOPS/mm² | 0.013 TOPS/W | 132 | √ | **X** |
| Feldmann et al [28] | N/A | N/A | 64 | √ | √ |
| Fang et al [19] | 0.23 TOPS/mm² | N/A | N/A | √ | √ |
| Dong et al [20] | N/A | 121.7 pJ/OP | N/A | √ | **X** |
| Tianjic [62] | 84.08 GOPS/mm² | 1278 GOPS/W | N/A | N/A | √ |
| NVIDIA H100 [63] | N/A | 0.15 TOPS/W | N/A | N/A | **X** |

**Conclusion.** In summary, we present a programmable incoherent photonic neuromorphic computing chip equipped with 272 trainable model parameters, which supports the full deployment of a SNN layer. A core strength of this chip lies in its capability to execute both linear and nonlinear spike computations directly within the optical domain, eliminating the need for frequent optical-to-electrical conversion and thereby laying the foundation for high-efficiency computing.

Our design incorporates two specialized components with notable performance merits. First, the photonic synapse chip benefits from a simplified architecture that delivers low loss, low power consumption, and a compact footprint. Second, the photonic spiking neuron array chip achieves a low lasing threshold and small form factor through an optimized epitaxial wafer structure.

To address the inherent challenges of SNN training and enhance the accuracy and robustness of photonic SNNs, we further developed a software-hardware collaborative training-inference framework. This framework integrates software pre-training, photonic hardware in-situ training, and hardware-aware software fine-tuning, and this approach is generalizable to all photonic SNN hardware architectures. By introducing DCT for input data dimension reduction, we realized a high-performance lightweight frequency-domain SNN operating at a single time step (T=1). Hardware testing validated its effectiveness, achieving 90% accuracy on the MNIST dataset and 80.5% accuracy on the Fashion-MNIST dataset.

Compared with state-of-the-art PNN chips capable of supporting optical nonlinear computation, our work exhibits distinct advantages including high energy efficiency (1.39 TOPS/W for linear computations and 987.65 GOPS/W for nonlinear computations), high computing density (0.13 TOPS/mm$^2$ for linear computations and 533.33 GOPS/mm$^2$ for nonlinear computations), and low latency (320 ps). Overall, our research addresses two critical challenges including the lack of large-scale photonic nonlinear spike computation capabilities and the difficulty of training photonic SNNs. This work thus paves a promising path for the development of fully functional photonic SNN chips.

## Acknowledgements

We are grateful for financial supports from the National Natural Science Foundation of China (No.62535015, 62575231); The Fundamental Research Funds for the Centrale Universities (QTZX23041).

## Author contributions

S. Y. Xiang and Y. H. Zhang are co-first authors of the article. S. Y. Xiang proposed the idea, S. Y. Xiang, Y. H. Zhang prepared the manuscript, S. X. Shi, H. W. Zhao, D. Z. Zheng performed system experiments, Y. H. Zhang, X. X. Guo tested the chips, Y. N. Han, Y. Tian, L. Y. Zhang developed the algorithms, Y. C. Shi fabricated the samples, S. Y. Xiang, Y. Hao supervised the overall projects. All the authors analyzed and discussed the results.

## Competing interests

The authors declare no competing financial interests.

## Data availability.

The example dataset used in this paper are publicly available and can be accessed at https://yann.lecun.com/exdb/mnist/ for MNIST dataset, and https://github.com/zalandoresearch/fashion-mnist for fashion-MNIST dataset. Other data that support the findings of this study are available from the corresponding author upon reasonable request.